# Epitaxial growth and structural characterization of Pb(Fe$_{1/2}$Nb$_{1/2}$)O$_3$ thin films

W. Peng[a], N. Lemée[a,*], J. Holc[b], M. Kosec[b], R. Blinc[b], M. G. Karkut[a]

[a] *Laboratoire de Physique de la Matière Condensée, Université de Picardie Jules Verne, 33 rue Saint Leu, 80039 Amiens, France*

[b] *Jožef Stefan Institute, Jamova 39, 1000 Ljubljana, Slovenia*

We have grown lead iron niobate thin films with composition Pb(Fe$_{1/2}$Nb$_{1/2}$)O$_3$ (PFN) on (001) SrTiO$_3$ substrates by pulsed laser deposition. The influence of the deposition conditions on the phase purity was studied. Due to similar thermodynamic stability spaces, a pyrochlore phase often coexists with the PFN perovskite phase. By optimizing the kinetic parameters, we succeeded in identifying a deposition window which resulted in epitaxial perovskite-phase PFN thin films with no identifiable trace of impurity phases appearing in the x-ray diffractograms. PFN films having thicknesses between 20 nm and 200 nm were smooth and epitaxially oriented with the substrate and as demonstrated by RHEED streaks which were aligned with the substrate axes. X-ray diffraction showed that the films were completely *c*-axis oriented and of excellent crystalline quality with low mosaicity (x-ray rocking curve FWHM $\leq 0.09°$). The surface roughness of thin films was also investigated by atomic force microscopy. The root-mean-square roughness varies between 0.9 nm for 50-nm-thick films to 16 nm for 100-nm-thick films. We also observe a correlation between grain size, surface roughness and film thickness.

* Author to whom correspondence should be addressed: nathalie.lemee@u-picardie.fr

1. Introduction

Complex perovskite-type magnetoelectric multiferroics with the general formula $A(B^{I}B^{II})O_3$ have been widely investigated due to attractive multiple physical coupling phenomena and potential applications [1,2]. Among these compounds is lead iron niobate $Pb(Fe_{1/2}Nb_{1/2})O_3$ (PFN), a promising multiferroic material which was first discovered fifty years ago by Smolenskii *et al* [3]. It undergoes a paraelectric to ferroelectric transition at the Curie temperature $T_C \approx 385$ K [3] and a paramagnetic to antiferromagnetic transition at the Néel temperature $T_N \approx 145$ K [4,5]. Thus PFN is a ferroelectromagnet below the Néel temperature where ferroelectric and magnetic ordering coexist. Although the magnetoelectric (ME) effects in the vicinities of $T_C$ and $T_N$ have been reported [6-9], the exact nature of the ME coupling in PFN is still open. Obstacles to understanding these materials seem to be related to sample imperfections and/or inhomogeneities [10]. For bulk crystals and ceramics, there is disagreement about the crystal structure of PFN at room temperature: both monoclinic [10] and rhombohedral [11] structures have been proposed. Nonepitaxial PFN thin films, showing inferior properties to those of the bulk, have been fabricated by the sol-gel method [12], magnetron sputtering [13] and by pulsed laser deposition (PLD) [6,14]. Yan *et al.* [15] have reported significant improvements in the dielectric constant and maximum polarization in epitaxially grown PFN films. Due to the high volatility of lead, the epitaxial growth of stoichiometric PFN thin films without other secondary phases still presents a challenge. In this paper, we investigate the phase evolution of PFN according to the film deposition conditions in order to find an optimized deposition window for epitaxial growth. We will present the structural and microstructural properties of single phase PFN thin films.

2. Experimental

The PFN thin films were grown on single crystal (001) $SrTiO_3$ (STO) substrates by PLD, which is equipped with a 15 kV reflection high-energy electron diffraction (RHEED) system.

The beam of a KrF excimer laser (Lamda Physik, pulse duration 20 ns, $\lambda$ = 248 nm) was focused on a rotating stoichiometric ceramic target at 45º incidence. After deposition the films were immediately cooled to room temperature at the oxygen deposition pressure and the surface quality of the layers as well as the in-plane epitaxial relations of the film with respect to the substrate were checked in vacuum using RHEED. The deposition conditions including substrate temperature ($T_S$), the oxygen partial pressure ($P_{O2}$) and the laser parameters (pulse repetition rate, pulse energy density) were optimized so as to maximize the phase purity and the crystalline quality. The thicknesses were calculated according to deposition rates determined by x-ray Laue oscillations or by the Scherrer formula. The phase and crystal structure of thin films were analyzed using standard $\theta$-$2\theta$ x-ray diffraction (XRD) with Cu K$\alpha$ radiation (Siemens D5000). The out-of-plane mosaicity and in-plane epitaxy investigations were performed using four-circle XRD (Scintag PAD V) operated in $\omega$-scan and $\phi$-scan modes, respectively. The thin film surface morphology was determined with atomic force microscopy (AFM) operating in contact mode.

## 3. Results and discussion

In general, the thermodynamic stabilities of the perovskite and pyrochlore phases of Pb-based oxides are relatively similar, which can easily cause undesirable parasitic phases at temperatures favorable for the formation of the perovskite phase [16]. First, a coarse temperature window for PFN growth was chosen in terms of phase evolution and reasonable deposition rates, although the perovskite and other parasitic phases coexist. Then, the deposition temperature was further optimized so as to obtain the perovskite phase while other kinetic parameters were adjusted to provide appropriate supersaturation and diffusion for epitaxial growth. We found that PFN thin films with high crystalline quality only can be achieved in a narrow deposition window with the optimized conditions of $T_S$ = 670 ºC ( $\Delta T_S \approx$ ±20 ºC), $P_{O2}$=0.2 mbar, and laser energy density of 1.5 J/cm$^2$ with 2 Hz repetition rate. Fig.1 shows XRD $\theta$–$2\theta$ patterns of PFN films grown at 0.2 mbar between $T_s$ = 600 and 700 °C. At 600 °C, the diffraction pattern displays numerous peaks characteristic of the presence of (00$l$) PFN perovskite phase, Pb-rich $Pb_3O_4$ phases and (400)-oriented pyrochlore phase. When the deposition temperature is raised to 650 °C, the growth of the perovskite phase improves and the parasitic phase $Pb_3O_4$ is suppressed, but a small amount of pyrochlore phase still exists. Under the optimized deposition conditions, the diffraction pattern indicates that the PFN thin film exhibits completely $c$-axis orientation without any trace of pyrochlore phase or other parasitic phases in the limit of the sensitivity of the diffractometer. As the deposition temperature is further increased, additional diffraction peaks appear at 29.76° and 61.4° to which we have assigned the (222) and (444) peaks of the pyrochlore phase, respectively. This is based on our observations of the pyrochlore ($hhh$), ($h$00) peak positions in 500-nm-thick PFN films. These results indicate that the reappearance of the pyrochlore phase at higher deposition temperatures is due to lead deficiency. This tendency is often observed for Pb-based complex perovskite compounds, where Pb deficiency due to lead volatilization or re-evaporation is attributed to promote the formation of the pyrochlore phase [17-20].

Figure 2 presents XRD $\theta$–$2\theta$ patterns for PFN films with different thickness between 20 nm and 300 nm grown at 670 °C and 0.2 mbar. It was found that the pyrochlore phase begins to reappear for the 300-nm-thick film even under the optimized conditions. This is probably a signature of excessive lead loss in the stoichiometric target induced by the repetitious laser ablation of the target. Indeed, we have found this to be the case for a $0.8Pb(Fe_{1/2}Nb_{1/2})O_3$-$0.2Pb(Mg_{1/2}W_{1/2})O_3$ ceramic target on which we used energy dispersive spectroscopy (EDS) [21] to perform stoichiometric analysis after 2000 laser pulses and 5000 laser pulses. The lead content decreased from 51 % for the former and 66 % for the latter number of pulses. It is not unreasonable to believe that similar behavior occurs for a pure PFN target. In particular the Pb-enriched ceramic targets have been already utilized successfully to compensate for the high volatility of Pb and prevent Pb deficiency in films grown by PLD [17,18,22-24].

For the single phase PFN films, the in-plane biaxial compressive strains are expected due to the lattice mismatch between film and substrate (~ 2.7 % for STO), which should elongate the out-of-plane lattice parameters for epitaxial PFN films. We calculated the out-of-plane lattice parameters using several (00$l$) Bragg peaks so as to improve the accuracy and found that, in the thickness range between 20 and 200 nm, the lattice parameters decrease monotonically from 0.4089 to 0.406 nm, which is slightly larger than that of the PFN bulk value (0.4012 nm). Although a large in-plane lattice mismatch exists for this heteroepitaxial system, Laue oscillations were still observed up to 50 nm (see Fig.3). It should be mentioned that the thickness deduced from the Laue oscillations, in these films with thickness below 50 nm, is identical to that deduced from the Scherrer formula, implying that the out-of-plane order is high and coherent over the film thickness.

In order to characterize the crystalline quality, we performed XRD $\omega$-scan of the PFN (001) reflection as shown in Fig. 4. The narrow rocking curves suggest that the films have high crystalline quality with low mosaicity. Indeed, for 20-nm-thick PFN thin film, the full-width at half-maximum (FWHM) is very close to that of the substrate peak. The in-plane orientation between the PFN film and the STO substrate was investigated by XRD $\phi$-scan, as shown in Fig.5a. Four STO {220} diffraction peaks are spaced by 90° and four PFN {220} peaks with the same spacing are also found at the same angular positions. In addition, Fig.5b illustrates the RHEED patterns for a PFN film taken at azimuthal incidence along the STO [100], [110] and [120] directions. The RHEED patterns between the film and substrate (not shown here) are similar. From the results of the XRD $\phi$-scan and RHEED azimuthal analysis, we establish the following epitaxial relationship of the films with the substrate: PFN(001)[100] ∥ STO(001)[100]. Note that the pseudocubic notation is used here for PFN.

The sharp RHEED streaks indicate the smooth surfaces of the films. This is also confirmed by an AFM morphology investigation. In Fig. 6, typical AFM images of PFN thin films with thicknesses of 50 and 100 nm are presented. A well crystallized and very smooth surface was achieved in the PFN thin films. The root-mean-square (RMS) roughness is merely 0.9 nm over the $2 \times 2$ $\mu m^2$ scan area for the 50-nm-thick PFN film. For the thicker 100 nm film, the surface becomes rougher with a RMS value of about 16 nm. We note that the grain sizes of the films simultaneously increase from 78 nm (50 nm thick) to 110 nm (100 nm thick), indicating a correlation between grain size, surface roughness and film thickness.

## 4. Conclusions

The deposition parameters for epitaxial PFN grown on (001) STO substrates by pulsed laser deposition have been explored in terms of thermodynamic and kinetic processes. It was found that the growth of PFN thin films is only possible in a narrow temperature window. At higher temperature, the pyrochlore phase is detected while at lower temperature an additional parasitic phase $Pb_3O_4$ is strongly favored. For thicknesses below 200 nm, perovskite-phase PFN epitaxial thin films have been successfully obtained under optimized deposition conditions. As a result, the PFN thin films exhibit high crystalline quality and smooth surfaces, providing additional opportunities to further investigate magnetoelectric coupling and local ferroelectric/magnetic properties.

**Acknowledgments**

We thank Dr. Olivier Durand-Drouhin for assistance with the AFM measurements. The authors are grateful to the Region of Picardie for financial support. This work was partially supported by the European 6[th] Framework STREP: "MULTICERAL" (Grant no. FP-6-NMP-CT-2006–032616).

**Figure captions**

Fig.1. XRD $\theta$–$2\theta$ scans for PFN thin films deposition at 0.2 mbar and different temperatures. Vertical dashed lines correspond to the position of the diffraction peaks of PFN. The peaks are indexed with the following convention: P: (00*l*) PFN, S: (00*l*) STO, ♦: (*h*00) $Pb_3O_4$, *: (*hhh*) pyrochlore, #: (*h*00) pyrochlore.

Fig.2. XRD $\theta$–$2\theta$ scans for various thickness PFN films grown under optimized deposition conditions. The same convention as in Fig.1 is adopted.

Fig.3. (Color online) XRD Laue oscillations (solid line) and simulation (dashed line) around the (001) Bragg peak for a 50-nm-thick PFN thin film.

Fig.4. XRD rocking curves of (001) Bragg peaks for epitaxial PFN thin films with (a) 20 nm and (b) 200 nm thickness.

Fig.5. In-plane epitaxy of 200-nm-thick PFN thin films. (a) XRD $\phi$-scans of {220} reflection planes for PFN and STO. (b) RHEED patterns of PFN films with different azimuthal incidence.

Fig.6. AFM images for epitaxial PFN thin films with (a) 50 nm and (b) 100 nm thickness.

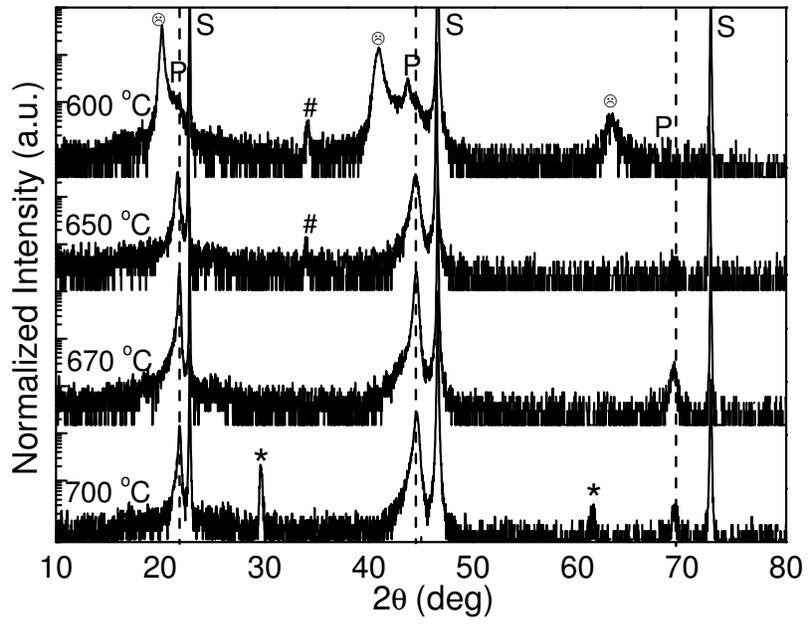

Fig.1

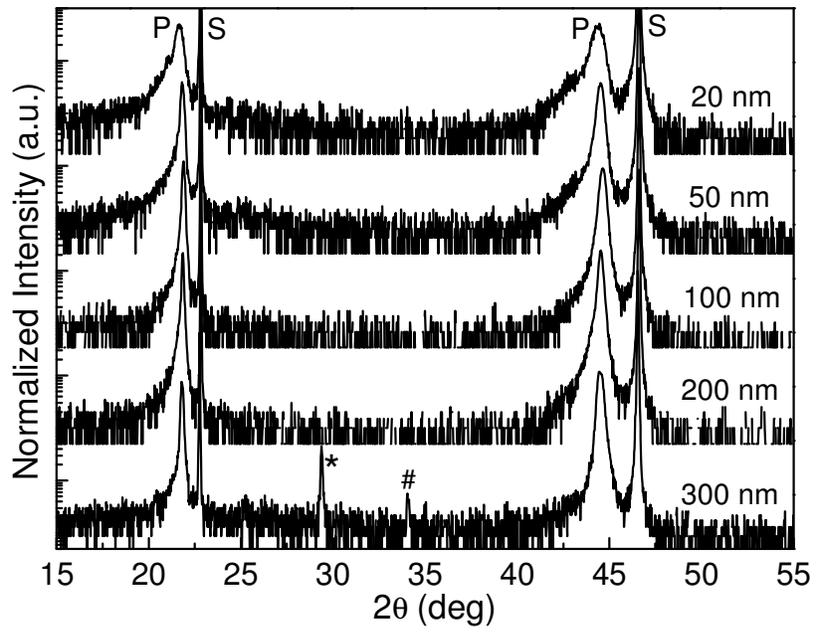

Fig.2

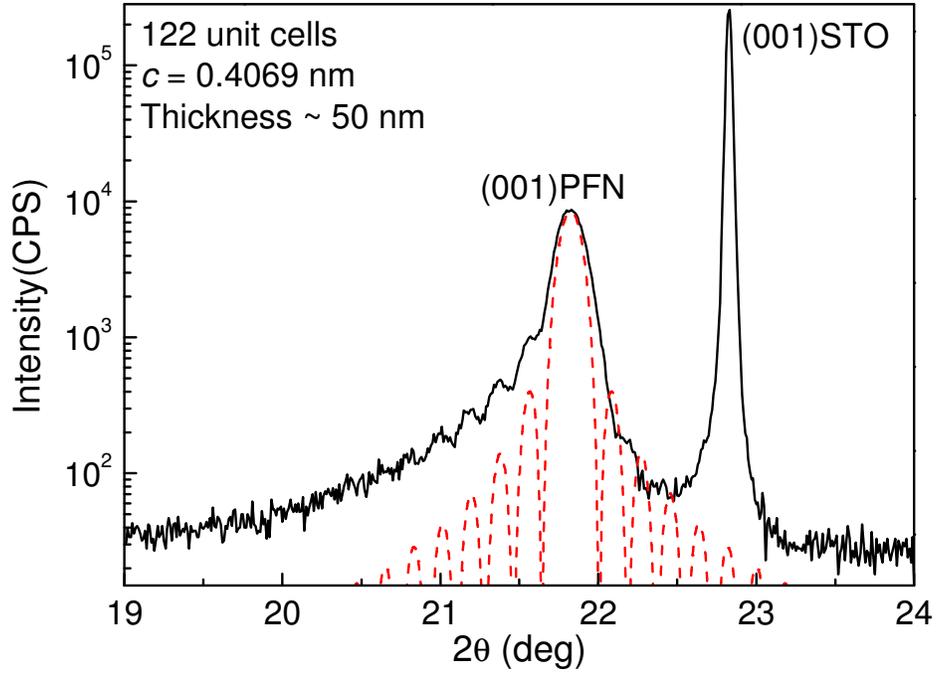

Fig.3

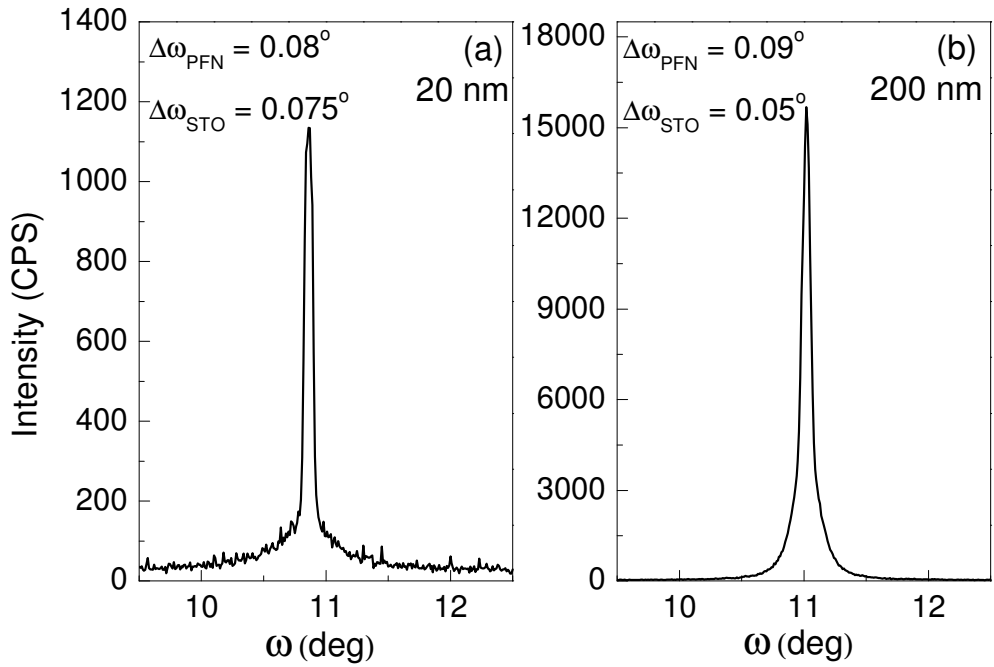

Fig. 4

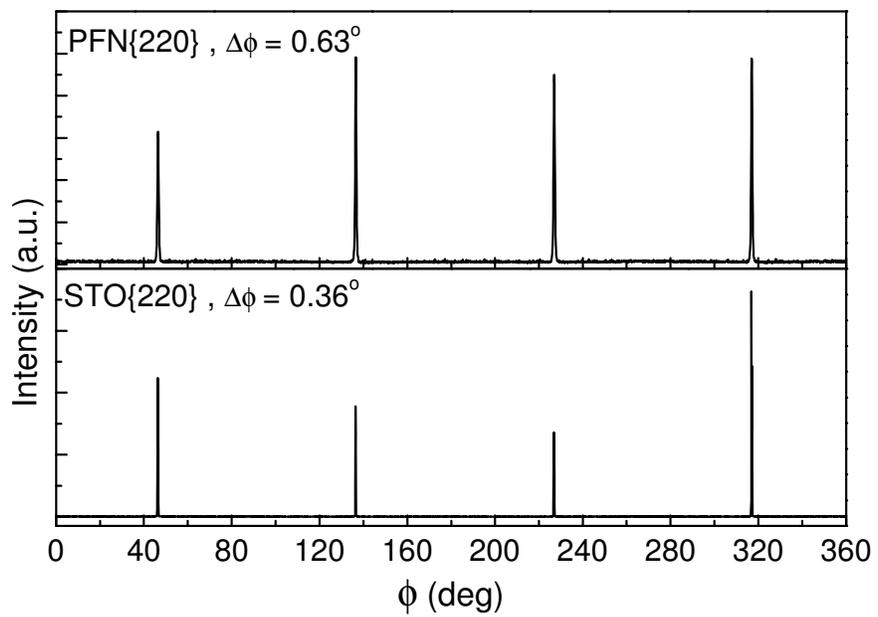

Fig. 5a

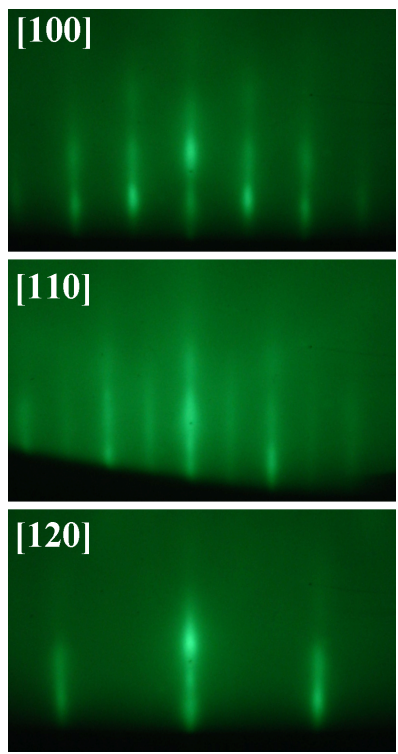

Fig. 5b

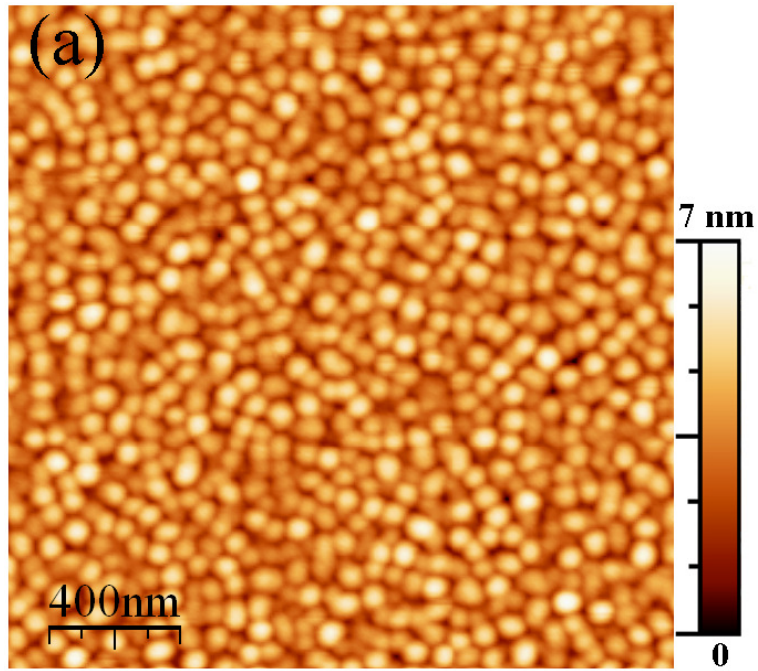

Fig. 6a

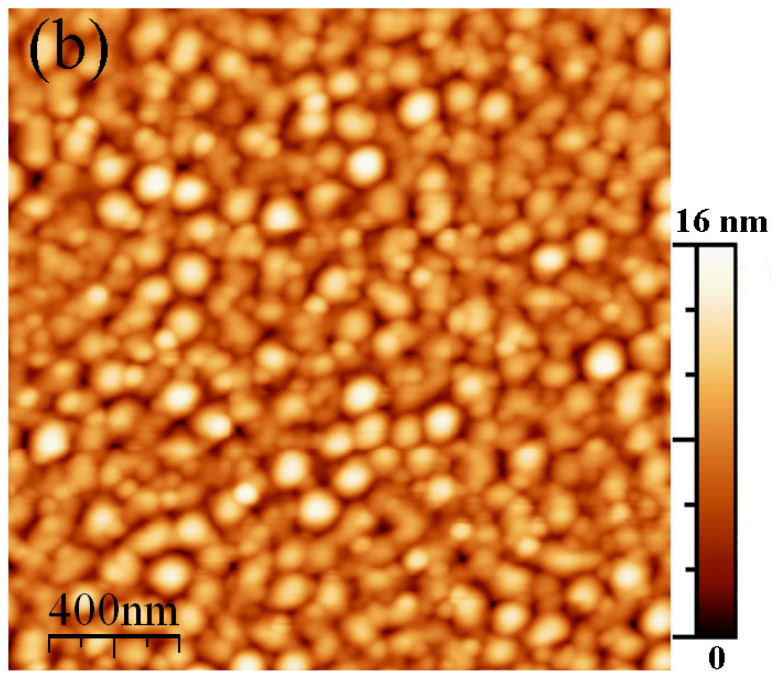

Fig. 6b